\documentclass[epj]{svjour}
\usepackage{amssymb}
\usepackage{amsmath}
\usepackage{graphicx}

\begin{document}

\title{Stochastic dynamics of the magneto-optical trap}
\author{Daniel Hennequin}
\institute{Laboratoire de Physique des Lasers, Atomes et Mol\'{e}cules,\\
Unit\'{e} mixte du Centre National de la Recherche Scientifique,\\
Centre d'Etudes et de Recherches Lasers et Applications, \\
B\^{a}t. P5, Universit\'{e} des Sciences et Technologies de Lille,\\
F-59655 Villeneuve d'Ascq cedex - France}
\date{Received: date / Revised version: date}

\abstract{The cloud of cold atoms obtained from a magneto-optical trap is known to
exhibit two types of instabilities in the regime of high atomic densities:
stochastic instabilities and deterministic instabilities. In the present
paper, the experimentally observed stochastic dynamics is described
extensively. It is shown that it exists a variety of dynamical behaviors,
which differ by the frequency components appearing in the dynamics. Indeed,
some instabilities exhibit only low frequency components, while in other
cases, a second time scale, corresponding to a higher frequency, appears in
the motion of the center of mass of the cloud. A one-dimensional stochastic
model taking into account the shadow effect is shown to be able to reproduce
the experimental behavior, linking the existence of instabilities to folded
stationary solutions where noise response is enhanced. The different types
of regimes are explained by the existence of a relaxation frequency, which
in some conditions is excited by noise.
}
\maketitle

\section{Introduction}

\begin{figure}[tph]
\centerline{\resizebox{0.45\textwidth}{!}{\includegraphics{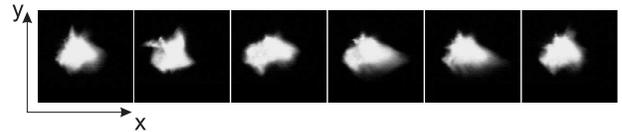}}}
\caption{Sequence of snapshots showing the time evolution of the unstable
atomic cloud. Snapshots are presented in the chronological order, each one
being separated by 120 ms.}
\label{fig:film}
\end{figure}

Magneto-optical traps (MOT) produce clouds of cooled atoms at temperature as
low as the $\mu $K. The achievement of such clouds opened many perspectives,
not only in the field of fundamental atomic physics, as e.g. in the domain
of the atomic dynamics or the quantum chaos\cite{QC}, but also leads to
several potential applications, in particular the improvement by several
orders of magnitude of the atomic clock\cite{clocks}. The MOT is also the
first stage to produce lower temperatures, in particular to obtain
Bose-Einstein condensates \cite{nobel}. Although the use of MOTs is
relatively well mastered, some details of the experimental setup remain
empirical, because of the existence in the cloud of instabilities that are
not well understood. Some studies showed that when the trapping beams are
misaligned, the cloud may be spatially altered and become unstable. In
particular, ring-shaped clouds and chaos have been observed, and attributed
to a vertex force \cite{shadow,mis}. However, the main instabilities
encountered in the experiments concern well-aligned MOTs. In that case, the
cloud, which is usually more or less ellipsoidal, has a complex irregular
shape, with an inhomogeneous atomic density (fig \ref{fig:film}). This shape
may not be stable, and changes as a function of time: a typical example of
these instabilities is shown in fig. \ref{fig:film}.

There is a deep interest to identify the nature of these instabilities, in
the aim to control them, and possibly to take advantage of them. To
illustrate these points, let us remember that instabilities may originate in
many mechanisms, which can be classified in two families: stochastic or
deterministic. In the latter, called also deterministic chaos, the dynamics
is described by a set of deterministic equations. When the number of
equations is small (low-dimensional deterministic chaos), it has been shown
that the dynamics can be controlled to reach various states that are not
accessible otherwise, as unstable states\cite{controlstab} or periodic
behaviors\cite{controlper}. It is also well known that the description of
such a dynamics gives access to numerous parameters, which are not
measurable when the behavior is stationary\cite{paramaccess}. The cost for
these informations is that there is no other way to reduce the
instabilities. On the contrary, if the origin of the instabilities is
stochastic, i.e. due to external noise, a reduction of instabilities is
obtained by reducing the noise, but the instabilities have no physical
meaning, in that sense that they are not intrinsic to the physics of the
system.

In \cite{nousprl}, a first type of instabilities, arising for low beam
intensities, has been depicted. It was showed that these so-called
``stochastic instabilities'' are induced by the absorption of the trapping
beams in the cloud. A model taking into account this so-called shadow effect
showed that from a dynamical point of view, instabilities arise through a
stochastic resonancelike phenomenon, namely the coherent resonance, linked
to a Hopf bifurcation in the stationary solutions of the MOT. It is well
known that such a bifurcation usually leads to periodic instabilities, and
indeed, a recent study evidenced such instabilities, purely deterministic 
\cite{InstDet}. However, a correct description of these self-oscillations
required to modify the model in \cite{nousprl}, in particular to take into
account the spatial distribution of the cloud.

In the present paper, the results presented in \cite{nousprl} are detailed,
and extended to a larger range of parameters, where new regimes appear, in
particular a behavior without resonance frequency, contrary to the
instabilities described in \cite{nousprl}. Then the modified model
introduced in \cite{InstDet} is accurately described, and the relative
domains of appearance of stochastic and deterministic instabilities are
discussed. The difference appears to be the result of the preeminent role of
noise in some parameter range, in particular in the vicinity of the
bifurcations. The stochastic instabilities predicted by the model are
studied, and they are shown to be in agreement with those experimentally
observed. In particular, an interpretation \ of the two types of stochastic
instabilities, with and without resonance frequency, is found.

The paper is organized as follows. After this introduction, the section \ref
{S2ExpSetup} describes the experimental setup. Then, a detailed analysis of
the experimental observations of the stochastic instabilities is presented
(section \ref{S3ExpRes}), showing in particular the existence of two types
of instabilities. Section \ref{S4DetMod} is devoted to the construction of a
simple 1D model, already presented in \cite{InstDet} in a less detailed way.
In section \ref{S5StatSol}, the stationary solutions of the model are
discussed. Finally, in section \ref{S7StaFold}, the noise induced dynamics
instabilities is studied and compared to the experimental observations.

\section{Experimental set-up}

\label{S2ExpSetup}We work with a Cesium-atom MOT in the usual $\sigma
_{+}-\sigma _{-}$\ configuration, with three arms of two counter-propagating
beams obtained from the same laser diode. The waist $w_{T}$ of the trap
beams may be varied from typically $3$ to $10$~mm. Two configurations are
possible: in the first one, all six beams are independent, by opposition to
the second configuration, where counter-propagating beams result from the
reflection of the three forward beams. In the last case, the intensity
asymmetry resulting from the absorption of the forward beam by the cloud,
generates a center-of-mass motion, while in the first case, instabilities
are characterized by symmetrical bursts on the cloud shape, much more
difficult to measure. However, as the nonlinearities involved in both cases
are the same, we expect that the dynamics will be fundamentally of the same
nature, and thus we choose the configuration with retro-reflected beams.

A full description of the unstable dynamics of the atomic cloud will be
presented in the next section. However, to make easier the understanding of
this paragraph, let us depict them briefly. As shown in the introduction
(fig. \ref{fig:film}), instabilities consist in a deformation of the spatial
atomic distribution, leading to fluctuations of the shape of the cloud.
Therefore, the relevant {\em dynamical variables} allowing us to describe
instabilities, could be the shape of the cloud (i.e. for example the {\em %
local} velocities and atomic densities in the cloud). This type of
description corresponds to a high dimensional model, associated with partial
differential equations. Here, for the sake of simplicity, we choose to limit
our description to the center of mass (CM) location ${\bf r}$, and the {\em %
total} number of atoms $n$ in the atomic cloud. This allows us to model the
system with ordinary differential equations, and reduces the dimension to
seven, and even three in a 1D model. As it is shown in the following, the
use of this description appears to be sufficient to understand the main
mechanisms of the instabilities.

To measure ${\bf r}$ and $n$, we used two 4-quadrant photodiodes (4QP)
forming an orthogonal dihedral and measuring the fluorescence of the cloud.
The differential signal of the 4QP allows us to monitor the motion of the
CM. Using only two 4QPs does not allow us to reconstruct the actual 3D
motion, but gives access to projections of this motion on two different
axes. This prevents the measure from line-of-sight effects due to the
optical thickness of the cloud. We checked that whatever the type of
dynamical behavior, the motion components $r$ recorded by both 4QP have the
same properties and are qualitatively identical. In the following, the ``CM
motion'' refers to a component $r$ of this motion recorded by one of the
4QPs. The second dynamical variable, namely the number of atoms inside the
cloud, is deduced from the total signal received from the 4QPs. In addition
to the 4QPs, two video cameras monitor the shape of the cloud. Because of
their poor resolution as compared with the 4QPs, the video cameras are not
used to record the dynamical variables. They have been essentially used in
the first stages of the experiment, to control that there is no discrepancy
between the shape dynamics and the CM dynamics.

Instabilities depend on the MOT parameters. Among these parameters, some are
easily controllable, and will be referred in the following as {\em control
parameters}. They are the detuning $\Delta _{0}$ of the MOT, the magnetic
field gradient $G$, the MOT beam intensities $I_{1}$ and the repumper laser
intensity $I_{rep}$. Other parameters cannot be considered as control
parameters, because they are not easily controllable or measurable in our
experimental setup. Among these parameters, the alignment of the MOT beams
appeared to be crucial in the experiment. We limit the present study to the
case where beams are aligned. When misaligned, most of the dynamical
characteristics of the cloud change qualitatively \cite{mis}. The MOT beam
waists play also a main role on the dynamics, but in practice, they cannot
be changed easily independently from the other parameters, in particular $%
I_{1}$. However, two different values have been used in the experiments, and
their effects on the dynamics are well understood. Finally,\ the vapor
pressure in the cell has probably also a large influence on the dynamics.
Unfortunately, this parameter is not easily measurable in our experimental
setup. Moreover, as it is shown later, it does not appear explicitly in the
model, but its impact on the dynamics may be estimated through the
equilibrium population of the cloud. The parameter ranges explored in the
present experiment are summarized in Tab. \ref{tabexpparams}.

\begin{table}[h]
\caption{Range of the parameters used in the present experiment. $G$ is the
magnetic field gradient, $I_1$ is the intensity of the forward beam, $%
\Delta_0$ is the detuning and $w_T$ is the trap laser beam waist. $I_s$ is
the saturation intensity ( $I_s=1.1 $ mW) and $\Gamma$ is the natural width
of the transition. The last column indicates the default parameter values
used to obtain the results reported in the present paper.}
\label{tabexpparams}
\begin{tabular}{rcc}
& range & default set \\ \hline
$G$ (Gcm$^{-1}$) & $G\leq 14$ & 14 \\ 
$I_+=I/I_s$ & $4\leq I_+\leq 20$ & 6 \\ 
$w_T (mm)$ & $3\leq w_T\leq 10$ & 3 \\ 
$\Delta_0$ & $\Delta_0\leq -0.5$ & -
\end{tabular}
\end{table}

\section{Experimental results}

\label{S3ExpRes}Instabilities have been described in \cite{nousprl} for a
given set of parameters. In the following, we extend this description for
the whole range of parameters where stochastic instabilities appear. A first
fundamental control parameter is the MOT beam intensity, whose value
determines the type of observed instabilities. At low MOT beam intensities,
typically less than $10I_{S}$ ($I_{S}=1.1$~mW/cm$^{2}$ is the saturation
intensity), the cloud exhibits {\it S} instabilities ({\it S} stands for
Stochastic). For MOT beam intensities larger than $10I_{S}$, {\it C} (for
Cyclic) instabilities appear \cite{InstDet}. As the aim of the present paper
is to discuss about the stochastic instabilities, we keep a detailed
presentation of the deterministic {\it C} instabilities for another paper.
However, the two types of instabilities cannot be completely separated, and
the domains of appearance of both types of instabilities will be discussed.

\begin{figure}[tph]
\centerline{\resizebox{0.45\textwidth}{!}{\includegraphics{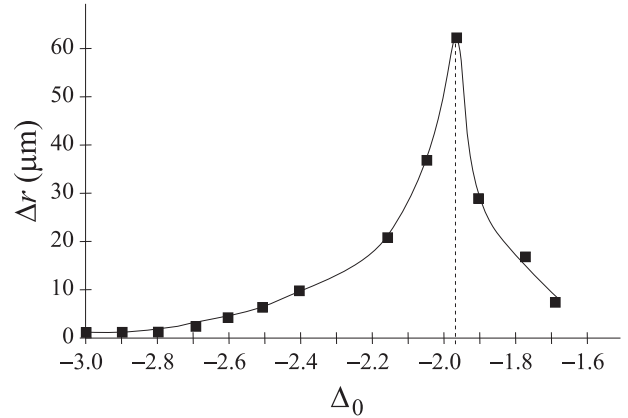}}}
\caption{Evolution of the signal amplitude $\Delta r$ as a function of the
detuning $\Delta_0$. Parameters are those of the default set of Tab. \ref
{tabexpparams}.}
\label{fig:expmax}
\end{figure}

As discussed above, {\it S} instabilities appear at low MOT beam
intensities, and are characterized by large fluctuations of the shape of the
cloud appearing in a limited range of the parameters. From the experimental
point of view, this last point is essential, because it is at the origin of
the introduction of the concept of instabilities of the MOT. Indeed, if the
noisy dynamics is the same whatever the parameter values, it is clear that
the problem becomes as trivial as the reduction of technical noise in an
experiment. On the contrary, the motion of the MOT grows suddenly in a
narrow range of the parameters, as it is illustrated in fig. \ref{fig:expmax}%
, where the amplitude $\Delta r$ of the $r$ fluctuations is represented as a
function of the control parameter $\Delta _{0}$. Far from the unstable area,
e.g. at large $\left| \Delta _{0}\right| $, the cloud is stable in shape and
density. When the control parameter is tuned to the unstable area,
instabilities appear progressively, and the amplitude grows until a maximum $%
\Delta r_{\max }$ at $\Delta _{0\max }\simeq -1.95$ ($\Delta _{0}$ is
expressed in units of the natural width $\Gamma $ of the atomic transition).
There is no abrupt boundary between the stable and unstable areas: the
limits given below corresponds to $\Delta r=\Delta r_{\max }/10$. In this
case, for the set of parameters of fig. \ref{fig:expmax}, instabilities
appear for $-2.5<\Delta _{0}<-1.7$ (for $\Delta _{0}>-1.7$, the population
in the cloud vanishes). The unstable range depends on the other parameters,
such as the beam intensities, the vapor pressure or the magnetic field
gradient. For example, for a less populated cloud, due e.g. to a different
vapor pressure in the cell, all other parameters being the same as
previously, instabilities will appear at smaller detuning. However, whatever
the parameters used in the experiments, in the range given in Tab. \ref
{tabexpparams}, we have $-3\lesssim \Delta _{0\max }\lesssim -1$, and
instabilities never occur on a range larger than $1$.

\begin{figure}[tph]
\centerline{\resizebox{0.45\textwidth}{!}{\includegraphics{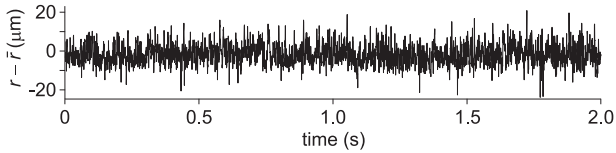}}}
\caption{Experimental record of the time evolution of a component of the CM
location of the atomic cloud. Experimental parameters are the default ones
given in Table \ref{tabexpparams}, with $\Delta_0=-2$. The {\em mean} cloud
population is $1.5\times 10^{8}$ atoms and the cloud size is 1 mm.}
\label{fig:exp1time}
\end{figure}

\begin{figure}[tph]
\centerline{\resizebox{0.45\textwidth}{!}{\includegraphics{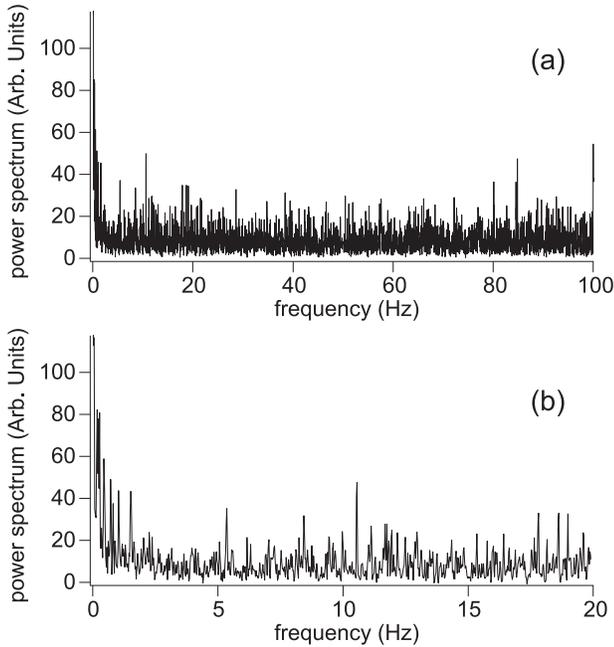}}}
\caption{Power spectra corresponding to the behavior illustrated in fig. \ref
{fig:exp1time}. In (a) CM location, and in (b) cloud population. The scales
are linear.}
\label{fig:exp1freq}
\end{figure}

Fig. \ref{fig:exp1time} shows a typical unstable behavior of $r$. It appears
as an erratic signal, with a flat spectrum (fig. \ref{fig:exp1freq}). Low
frequency significant components appear typically for values smaller than $%
\nu _{n}\approx 2$~Hz (fig. \ref{fig:exp1freq}b), and the dynamics is
essentially along the first bisector of the three forward beams. The
behavior of $r$ and $n$ are similar, with a cross correlation coefficient
larger than 0.8. To determine if these instabilities have a deterministic
origin, several tools are offered through the nonlinear dynamical analysis
of the time series. As the MOT is dissipative, a deterministic dynamics
should have an attractor, which can be reconstructed easily for a low
dimensional dynamics. The result, not presented here, appears as a set of
randomly distributed points: in particular, it does not present any fine
structure. Poincar\'{e} section and 1D maps confirm this absence of order in
the dynamics. This could be due to a lack of resolution of the measures, but
the general shape of the trajectories rather suggests that the behavior is
stochastic (or chaotic with a high dimensional dynamics). Because this
behavior appears to be a stochastic dynamics with only low frequency
components, it will be referred in the following as {\it S}$_{{\it L}}$
instabilities.

\begin{figure}[tph]
\centerline{\resizebox{0.45\textwidth}{!}{\includegraphics{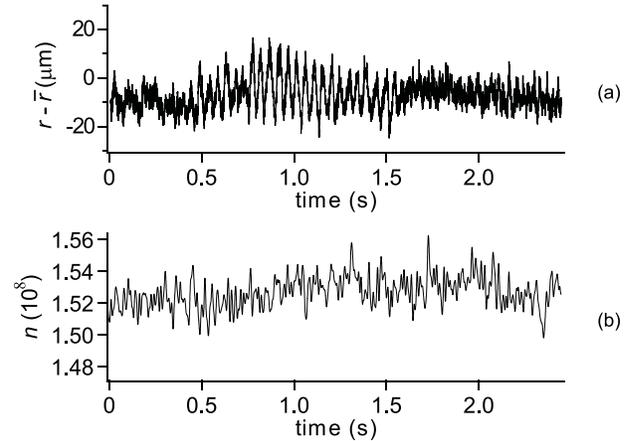}}}
\caption{Experimental record of the time evolution of a component of the CM
location of the atomic cloud in the case of the {\it {S}} behavior.
Experimental parameters correspond to the default set given in Table \ref
{tabexpparams}, with $\Delta_0=-1.5$. The {\em mean} cloud population is $%
1.5\times 10^{8}$ atoms and the cloud size is 1 mm. }
\label{fig:exp2time}
\end{figure}

\begin{figure}[tph]
\centerline{\resizebox{0.45\textwidth}{!}{\includegraphics{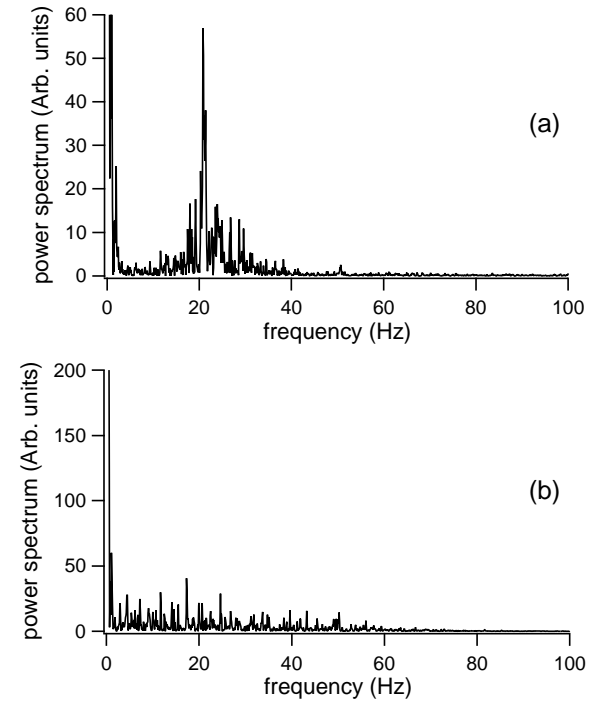}}}
\caption{Power spectra corresponding to the behavior illustrated in fig. \ref
{fig:exp2time}. In (a) CM location, and in (b) cloud population. The scales
are linear.}
\label{fig:exp2freq}
\end{figure}

In some situations corresponding to given ranges of parameters \cite{nousprl}%
, the ${\bf r}$ dynamics is altered by the appearance of spontaneous large
amplitude oscillation-like bursts. As illustrated in fig. \ref{fig:exp2time}%
, the signal inside the bursts is not periodic, although it is clearly
dominated by a given frequency. These bursts are relatively scarce, and the
global shape of the signal remains that of Fig. \ref{fig:exp1time}. The
bursts appear in fact as the most spectacular consequence of a deeper change
of the dynamics, which is the appearance of a second characteristic time.
This appears clearly in the spectrum of $r$ (Fig. \ref{fig:exp2freq}a) as a
peak centered at a frequency $\nu _{r}$, which depends on the parameter
values, but ranges typically between 10 and 100 Hz. The other
characteristics of the dynamics are not modified by the existence of bursts.
In particular, the spectrum still exhibits the low frequency component below 
$\nu _{n}$, and the bursts do not appear on the $n$\ behavior (Fig. \ref
{fig:exp2time}b and \ref{fig:exp2freq}b). The low frequency component
dynamics of $r$\ and the dynamics of $n$\ remain correlated. This behavior
will be referred in the following as {\it S}$_{{\it H}}$ instabilities.

When the trap beam intensity $I_{1}$ is increased, {\it S} instabilities
still exist, but they are progressively superseded by {\it C} instabilities.
These instabilities differ drastically from {\it S} ones\cite{InstDet}: they
can be either periodic or erratic, but in both cases, they are cyclic, and
the motion amplitude is much larger. However, as for {\it S} instabilities, 
{\it C} instabilities exist in a limited range of $\Delta _{0}$, which value
depends on the other parameters. But it is systematically between $\Delta
_{0}=-3$ and resonance, and the maximum detuning range is of the order of $1$%
.

\begin{figure}[tph]
\centerline{\resizebox{0.45\textwidth}{!}{\includegraphics{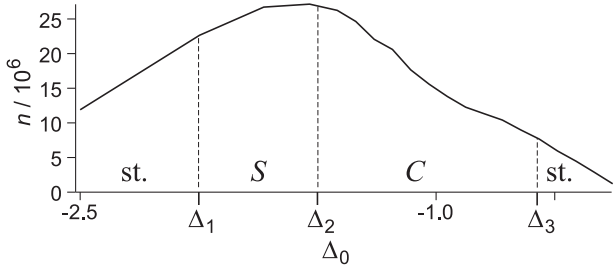}}}
\caption{This figure illustrates the evolution of the behavior as a function
of the detuning for $I_1=6.8$ and $I_{rep}=1.5$ mW/cm$^2$. The full line
reports the population, while the dashed lines separate the domain of
different behaviors: st. stands for stable, S for {\it S} instabilities and
C for {\it C} instabilities.}
\label{fig:VII3}
\end{figure}

As $I_{1}$ is increased, the disappearance of {\it S} instabilities occurs
progressively, in favor of {\it C} instabilities. For intermediate values of 
$I_{1}$, both types of instabilities exist. Their typical distribution
versus $\Delta _{0}$ is illustrated in Fig. \ref{fig:VII3}: far from
resonance, the cloud is stable; as the resonance is approached, {\it S}
instabilities appear for a detuning $\Delta _{0}=\Delta _{1}$. Then {\it C}
instabilities appear in $\Delta _{2}>\Delta _{1}$. If the detuning is still
increased, {\it C} instabilities disappear in $\Delta _{3}$ at the benefit
of a stable behavior. Finally, the cloud vanishes in $\Delta _{4}$. As $%
I_{1} $ is increased, the width $\delta _{12}=\Delta _{2}-\Delta _{1}$
decreases in favor of the interval $\delta _{23}=\Delta _{3}-\Delta _{2}$,
while the total unstable interval $\delta _{13}=\Delta _{3}-\Delta _{1}$
remains more or less constant. When {\it C} instabilities merge for $%
I_{1}=4I_{S}$, they appear on\ a narrow interval $\delta _{23}\gtrsim 0$.
This interval increases rapidly until $I_{1}=7.5I_{S}$ and $\delta _{23}=0.8$%
. For $I_{1}>7.5I_{S}$, $\delta _{23}$ increases more slowly, to reach the
value of $\delta _{23}=1$ in $I_{1}=20I_{S}$, where {\it S} instabilities
have completely disappeared.

To conclude this section, let us summarize the main properties of the
instabilities: they appear to be stochastic at a time scale larger than 0.5
s. They may exhibit a second dynamical component with time scale smaller
than $10^{-2}$ s and acting only on ${\bf r}$. The low frequency component
corresponds to a CM motion along the first bisector of the three forward
beams. Finally, they appear only in a limited range of the parameters,
suggesting a nonlinear origin of the instabilities. This last point was also
suggested by different tests showing that beam phase fluctuations are not at
the origin of instabilities \cite{nousprl}.

\section{Model}

\label{S4DetMod}To understand the dynamics observed in the experiments, we
build a phenomenological model taking into account the shadow effect. The
aim here is not to model as finely as possible the experimental system, but
on the contrary to make a model as simple as possible, enlightening the
fundamental mechanisms leading to the instabilities.

The model that we used here has already been described in \cite{InstDet}. It
is based on the shadow effect induced by the intensity gradients produced by
the absorption of the trapping laser beams in the cloud \cite
{shadow,dalibard}. The shadow effect generates a force compressing the
cloud. In the case where the counter-propagating beams result from the
reflection of the three forward beams, this compression is accompanied by a
displacement of the CM along the first bisector of the three forward beams:
indeed, the backward beams are less intense than the forward ones because of
the absorption in the cloud, and so the latter literally push the cloud off
its equilibrium location. The main role of the shadow effect in the
instabilities is suggested in the experiments (i) by the observed
correlation between the $n$\ and slow CM dynamics and (ii) by the fact that
the slow dynamical component of $r$\ displaces the cloud along the first
bisector of the three forward beams. The fundamental role of the shadow
effect in coupling the population and the CM position of cold atom clouds
was already reported in \cite{bistab}. In that experiment, bistability in
the population dynamics was observed by perturbing the atomic cloud with a
highly-focalized laser beam, and shadow effect is shown to be the dominant
nonlinearity.

We built a 1D model taking into account the shadow effect in a MOT where
counter-propagating beams result from the reflection of the three forward
beams. As we are interested here in the collective motion, we use as
variables the location $z$ of the center of mass of the cloud along the
unique axis $z$ of the system, and the number of atoms $n$ inside the cloud.
The origin of $z$\ coincides with the ``trap center'', that is, the zero of
the magnetic field. Thus, the motion of $z$ may be described by the
equation: 
\begin{equation}
\frac{d^{2}z}{dt^{2}}=\frac{1}{M}F_{T}
\end{equation}
where $M$\ is the mass of the cloud and $F_{T}$\ the global force exerted on
the atoms by the two counterpropagating beams. To evaluate $F_{T}$, we
assume a multiple scattering regime, i.e. a constant atomic density $\rho $
in the cloud. The atoms are distributed between $z_{\min }$ and $z_{\max }$,
and the quantity $\Delta z=z_{\max }-z_{\min }$ is the longitudinal size of
the cloud. We introduce a constant cross section $S$, allowing us to connect 
$\Delta z$ to $n$: 
\begin{equation}
n=\rho S\Delta z  \label{nvsdz}
\end{equation}
The repulsive force induced by the multiple scattering is an {\em internal}
force, and thus does not affect directly the center of mass motion. It is
also the case for the attractive force induced by the shadow effect, which
acts on $F_{T}$ only through the asymmetry between the forward and backward
beams. The forward beam is polarized $\sigma _{+}$ and has an input
intensity $I_{1}$. After the crossing of the cloud, its output intensity is $%
I_{2}$. This is also the input intensity of the backward beam, which is
polarized $\sigma _{-}$. After the crossing of the atomic cloud, the output
intensity of the backward beam is $I_{3}$. We have obviously $%
I_{1}>I_{2}>I_{3}$. $F_{T}$ is proportional to the number of photons
absorbed per unit time, i.e. $S(I_{1}-I_{2})/\hbar \omega _{L}$ for the
forward beam and $S(I_{2}-I_{3})/\hbar \omega _{L}$ for the backward beam ($%
\hbar \omega _{L}$ is the energy of one photon, $\omega _{L}$ being the
laser angular frequency). $F_{T}$ is obtained by multiplying the difference
between both quantities by $\hbar k_{L}$ 
\begin{equation}
F_{T}=\frac{S}{c}(I_{1}-2I_{2}+I_{3})  \label{newforce}
\end{equation}
To evaluate $I_{2}$ and $I_{3}$, we need to solve the propagation equation
of light inside the cloud. To simplify the calculations, we choose to model
a $F_{g}=0\rightarrow F_{e}=1$ transition: this is the simplest one leading
to a magneto-Doppler effect. This choice does not allow us to reproduce the
sub-Doppler effects, neither the photon redistribution between the two
waves, but as shown in \cite{InstDet}, it allows us to explain most of the
experimental behaviors.

The $z$ axis is chosen as quantification axis, and we introduce the ground
state $|g\rangle $ and the three Zeeman sub-levels of the excited state $%
|e_{0}\rangle $ et $|e_{\pm }\rangle $. Because of the kinetic momentum
conservation, $|e_{0}\rangle $ is not coupled to the light field, and thus,
the model is reduced to a three level system, $\{|g\rangle ,|e_{-}\rangle
,|e_{+}\rangle \}$. The interaction with light is governed by the Rabi
frequency $\Omega _{\pm }$ and the effective detunings $\Delta _{\pm }$,
taking into account the Doppler et Zeeman variations: 
\begin{equation}
\Delta _{\pm }=\frac{1}{\Gamma }(\Delta _{0}\mp kv\mp \omega _{B}^{\prime
}z)=\frac{1}{\Gamma }(\Delta _{0}\pm 2\delta )  \label{eq:delta}
\end{equation}
where $k$ is the wave vector of light and $\omega _{B}^{\prime }$ the Zeeman
shift, measured in angular frequency by unit of length. The value of the
atomic variables is given by the stationary solution of the optical Bloch
equations for the density matrix $\sigma $: 
\begin{equation}
\dot{\sigma}=\frac{1}{i\hbar }[H\;,\sigma \;]+\dot{\sigma}_{relax}
\label{blochPMO}
\end{equation}
The hamiltonian $H$ of coupling with light may be written as a function of $%
\Delta _{\pm }$ and $\Omega _{\pm }$ in the $\{|g\rangle ,|e_{-}\rangle
,|e_{+}\rangle \}$ basis: 
\begin{equation}
H=\hbar \left( 
\begin{array}{ccc}
0 & \Omega _{-}/2 & \Omega _{+}/2 \\ 
\Omega _{-}^{\ast }/2 & -\Delta _{-} & 0 \\ 
\Omega _{+}^{\ast }/2 & 0 & -\Delta _{+}
\end{array}
\right)
\end{equation}
The relaxation rates $\dot{\sigma}_{relax}$ are different for the excited
states, the optical coherences and the ground state: 
\begin{subequations}
\label{sigma}
\begin{eqnarray}
(\dot{\sigma _{ee}})_{relax} &=&-\Gamma \sigma _{ee}  \label{sigma1} \\
(\dot{\sigma _{eg}})_{relax} &=&-\frac{\Gamma }{2}\sigma _{eg}
\label{sigma2} \\
(\dot{\sigma _{gg}})_{relax} &=&-\Gamma (\sigma _{++}+\sigma _{--})
\label{sigma3}
\end{eqnarray}
\qquad The time scale of the internal variables is $\Gamma ^{-1}$, and so is
much shorter than that of the external variables $z$ and $v$. Thus the
stationary solution of Eq.~(\ref{blochPMO}) can be used. The solution for
the populations of the excited states $\Pi _{\pm }=\sigma _{\pm \pm }$ can
be obtained analytically. On the other hand, as the redistribution of
photons between the beams is forbidden, $\Pi _{\pm }$ is directly
proportional to the absorption rate of the beams, and the intensity gradient
can be written: 
\end{subequations}
\begin{equation}
\frac{dI_{\pm }}{dz}=\mp \Gamma \hbar \omega _{L}\rho \Pi _{\pm }
\label{eqpropint}
\end{equation}

By injecting the analytical solution of $\Pi _{\pm }$ in this equation, we
obtain the following equations of propagation: 
\begin{subequations}
\label{didz}
\begin{eqnarray}
\frac{dI_{+}}{dz} &=&-\Gamma \hbar \omega \rho \frac{\theta \omega
_{+}\omega _{-}+C_{-}\omega _{+}}{1-A_{+}\omega _{+}-A_{-}\omega
_{-}+3\theta \omega _{+}\omega _{-}}  \label{diplusdz} \\
\frac{dI_{-}}{dz} &=&\Gamma \hbar \omega \rho \frac{\theta \omega _{+}\omega
_{-}+C_{+}\omega _{-}}{1-A_{+}\omega _{+}-A_{-}\omega _{-}+3\theta \omega
_{+}\omega _{-}}  \label{dimoinsdz}
\end{eqnarray}
where: 
\end{subequations}
\begin{subequations}
\label{didzwhere}
\begin{eqnarray}
\omega _{\pm } &=&-\frac{I_{\pm }}{2\gamma } \\
\gamma &=&\frac{1}{4}\left[ \left( \alpha +\beta _{+}+\beta _{-}\right)
^{2}+\left( \frac{\beta _{+}-\beta _{-}}{\Delta _{0}}\right) ^{2}\right] \\
\alpha &=&\frac{1+2\mu }{4}+\Delta _{+}\Delta _{-}+\frac{\delta ^{2}}{2}%
\left( \mu -1\right) \\
\beta _{\pm } &=&\frac{\alpha }{2}\pm \left( \Delta _{0}\pm \delta \right)
\left( \frac{\delta }{2}+\Delta _{0}\left( \mu _{+}-\mu _{-}\right) \right)
\\
\mu _{\pm } &=&\frac{I_{\pm }}{4\left( 1+\delta ^{2}\right) } \\
\mu &=&\mu _{+}+\mu _{-} \\
\theta &=&\beta _{+}\beta _{-}+\alpha \left( \beta _{+}\mu _{+}+\beta
_{-}\mu _{-}\right) \\
A_{\pm } &=&2\beta _{\mp }+\alpha \left( 2\mu _{\pm }-\mu _{\mp }\right) \\
C_{\pm } &=&-\beta _{\pm }\pm \alpha \left( \mu _{+}-\mu _{-}\right)
\end{eqnarray}
These equations appear as the ratio of two polynomials of high order, and an
intuitive interpretation of this result is difficult. However, a numerical
resolution of these equations leads to a rigorous evaluation of $F_{T}$.

The cloud population dynamics is modeled by a ``feed-loss'' rate equation 
\cite{feedloss}: 
\end{subequations}
\begin{equation}
\frac{dn}{dt}=B\left( n_{e}-n\right)
\end{equation}
where we have introduce the population relaxation $B$ and the atom number in
the cloud at equilibrium $n_{e}$, which is linked to the loading rate $L$ by 
$L=Bn_{e}$. $n_{e}$ is assumed to depend on the CM location, to take into
account the losses variation when the cloud moves from the trap center. We
do not know the exact form of this dependence, because this variation may
have several origins. However, we can suppose that the main contribution
comes from the transverse distribution of the trap laser beam, which is
gaussian. For sake of simplicity, we keep only the first terms from its
Taylor's series, i.e. the quadratic term. In fact, we checked that the exact
dependence of $n_{e}$ does not change drastically the results given below.
So we write: 
\begin{equation}
n_{e}=n_{0}\left[ 1-\left( \frac{z}{z_{0}}\right) ^{2}\right]  \label{eq:ne}
\end{equation}
where $n_{0}$\ is the equilibrium cloud population at the trap center and $%
z_{0}$ a characteristic length. Atoms in $\left| z\right| >\left|
z_{0}\right| $ are considered as lost: this is taken into consideration when 
$n$ is deduced from Eq. \ref{nvsdz}.

\begin{figure}[tph]
\centerline{\resizebox{0.45\textwidth}{!}{\includegraphics{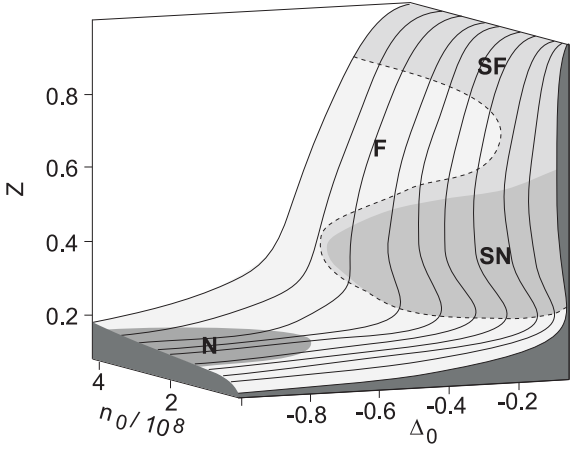}}}
\caption{Stationary solutions of equations \ref{eqred} versus $n_0$ and $%
\Delta_0$. The figure represents $Z_s$. Other parameters correspond to the
set \#1 given in table \ref{tabtheoparams}. The definition of the N, F, SN
and SF zones (each corresponding to different level of greys) is given in
the text. }
\label{fig:ZS}
\end{figure}

Finally, we introduce the reduced variables $Z=z/z_{0}$, $V=v/v_{r}$\ and $%
N=n/n_{0}$, where $v_{r}$ is the recoil velocity ($v_{r}=\hbar k/m$), and we
obtain the following autonomous system of equations: 
\begin{subequations}
\label{eqred}
\begin{eqnarray}
\frac{dZ}{dt} &=&V\frac{v_{r}}{z_{0}}  \label{eqred1} \\
\frac{dV}{dt} &=&\frac{1}{Mv_{r}}F_{T}  \label{eqred2} \\
\frac{dN}{dt} &=&B\left( 1-Z^{2}-N\right)  \label{eqred3}
\end{eqnarray}

Most of the theoretical parameters are the exact counterpart of the
experimental parameters, as e.g. the magnetic field gradient or the beam
intensities. In this case, we used in the model the same values as those of
Table \ref{tabexpparams}. It is not the case for all parameters, either
because of the simplicity of the model or because they cannot be measured
easily in the experiment. For example, $n_{0}$ has not a simple experimental
counterpart, but depends on several experimental parameters, as e.g. the
repumping laser intensity or the vapor pressure in the cell. For this
reason, a large interval of $n_{0}$ values has been used for the simulations
(Table \ref{tabtheoparams}). The density $\rho $ and the cross sectional
area $S$, which play the same role, have a meaning only in the context of a
1D model, while they correspond to variables in the experiments. They are
fixed in the simulations at experimental averaged values, and they have been
varied on a wide range to check that their value is not critical. Finally,
the parameter $z_{0}$ has not exact experimental counterpart, as it is
linked to both the trap beam waist and intensities. Indeed, the relevant
size for the atoms is not the beam waist, corresponding to an intensity
decreased by a ratio $e^{-2}$ as compared to the center of the beam, but
rather the location where the local beam intensity decreases under $I_{S}$.
This value is much larger than $w_{0}$ for intense beams.

\begin{table}[h]
\caption{Parameters used in the numerical simulations. The range corresponds
to the interval explored numerically, while the different sets refer to most
of the results presented in this paper.}
\label{tabtheoparams}
\begin{tabular}{rcccc}
& range & set \#1 & set \#2 &  \\ \hline
$G$ (Gcm$^{-1}$) & $14$ & $14$ & $14$ &  \\ 
$B$ (s$^{-1}$) & $3\leq B\leq 30$ & $3$ & $3$ &  \\ 
$I_1$ & $2\leq I_1\leq 30$ & 25 & 10 &  \\ 
$\rho$ (cm$^{-3}$) & $10^{10}\leq \rho\leq 3\times 10^{10}$ & $2\times
10^{10} $ & $2\times 10^{10}$ &  \\ 
$S$ (m$^2$) & $10^{-6}\leq S\leq 3\times 10^{-6}$ & $10^{-6}$ & $10^{-6}$ & 
\\ 
$z_0$ (m) & $10^{-2}\leq z_0\leq 3\times 10^{-1}$ & $3\times 10^{-2}$ & $%
3\times 10^{-2}$ &  \\ 
$n_0$ & $10^{7}\leq n_0\leq 10^{9}$ & $10^{8}$ & $2\times 10^{7}$ &  \\ 
$\Delta_0$ & $5\leq \Delta_0\leq 0$ & $-1.5$ & $-0.23$ & 
\end{tabular}
\end{table}

In order to check if the model can be more simplified, we tried to reproduce
the unstable dynamics of the cloud with the first terms of the Taylor's
series of Eqs.~(\ref{diplusdz}) et (\ref{dimoinsdz}), but we needed to keep
several orders and did not obtain simpler equations. Another possible
approximation concerns the different terms appearing in $\Delta _{\pm }$, in
Eq. \ref{eq:delta}. Using the values given in table II, one sees easily that
the $\Delta _{0}$ term and the Zeeman shift are of the same order of
magnitude, while the Doppler shift goes to zero at equilibrium, but can be
the largest term out of equilibrium. In these conditions, no approximation
is possible. Therefore, we use the equations (\ref{eqred}) in the
calculations.

To perform the comparison between the experiments and the present model, we
need to study the behavior of the system when noise is added. Noise can be
added in Eqs \ref{eqred} on any parameter, and as we were not able in our
experiments to identify clearly the main source of noise, we tried
theoretically several parameters, as the beam intensity $I_{1}$ or the
equilibrium population $n_{0}$. In this case, the stochastic model
corresponds to Eqs \ref{eqred} where the parameter $I_{1}$\ (resp. $n_{0}$)
is replaced by $\left( 1+\zeta \right) I_{1}$\ (resp. $\left( 1+\zeta
\right) n_{0}$), where $\zeta (t)$ is the noise component. We also simulated
several types of noise: gaussian white noise, but also colored noise with
different distributions of the frequencies. All configurations give
identical results: on the one hand, the choice of the noisy parameter is not
critical; on the other hand, the spectrum of noise does not alter the
response spectrum, except obviously that it depends on the relative weight
of the different frequencies in the noise. Thus, for sake of clarity, all
the results reported in the following have been obtained by applying with
gaussian white noise on $I_{1}$.

\section{Stationary solutions}

\label{S5StatSol}In the previous section, we built a relatively simple model where the cloud
is described by a set of three equations that do not depend explicitly on
the time. Such a system is known to be able to exhibit a complex dynamics,
including chaos, which could explain the experimental instabilities. To know
if such a complex dynamics occur in our conditions, the first step is to
evaluate the stability of the stationary solutions. The stationary solutions
($V_{s}$, $Z_{s}$, $N_{s}$) of the model with the shadow effect are easily
deduced from Eq.\ref{eqred} when the left side is put to zero. Eq. \ref
{eqred1} gives immediately $V_{s}=0$. From Eq. \ref{eqred3}, one finds that
the stationary solutions $Z_{s}$ of the CM location and $N_{s}$ of the
population are linked by the simple expression:

\end{subequations}
\begin{equation}
N_{s}=1-Z_{s}^{2}
\end{equation}
Therefore, the discussion is reduced to the analysis of $Z_{s}$. $Z_{s}$ is
the solution of $F_{T}=0$ in Eqs \ref{eqred2}. This equation can be resolved
numerically: its global shape is illustrated in Fig. \ref{fig:ZS}, where it
is plotted as a function of $\Delta _{0}$\ and $n_{0}$.

A first characteristic of the $Z_{s}$\ diagram is that $Z_{s}$ goes to $1$
(i.e. $N_{s}$ goes to zero) at resonance. This vanishing of the cloud, also
observed in the experiments, is a well-known consequence of the inefficiency
of the Doppler cooling close to resonance. It is here enhanced by the shadow
effect and the displacement of the cloud. The disappearance occurs suddenly
for small $n_{0}$, and becomes softer as $n_{0}$ increases. An interesting
point is that for small $n_{0}$, the abrupt increase of $Z_{s}$ is linked to
a very narrow bistable cycle. As $n_{0}$ is increased, the bistable cycle
shifts towards smaller $Z_{s}$ and smaller $\Delta _{0}$, and the vanishing
of $Z_{s}$ becomes progressive.

\begin{figure}[tph]
\centerline{\resizebox{0.45\textwidth}{!}{\includegraphics{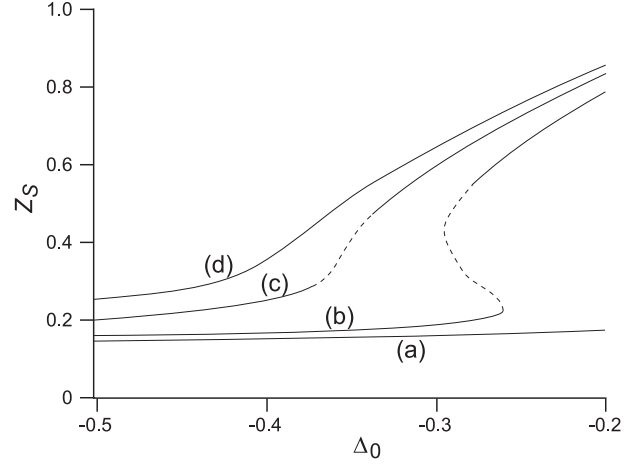}}}
\caption{Evolution as a function of the detuning of the stationary solution $%
Z_s$ of equations \ref{eqred}. The full (resp. dashed) line corresponds to a
stable (resp. unstable) solution. In (a), $n_0=0.5\times 10^8$; in (b) $%
n_0=2.5\times 10^8$; in (c), $n_0=3.4\times 10^8$; in (d), $n_0=4\times 10^8$
Other parameters correspond to the set \#1 of Table \ref{tabtheoparams}}
\label{fig:foldex}
\end{figure}

The main characteristic of the $Z_{s}$\ diagram is the presence of several
abrupt slope changes in the stationary solutions, leading to a fold in the
parameter space. The shape of the fold depends on the parameters, in
particular on $n_{0}$. Fig. \ref{fig:foldex} shows four examples
corresponding to situations leading to basically different atomic dynamics.
For $n_{0}=0.5\times 10^{8}$ (fig. \ref{fig:foldex}a), $Z_{s}$ increases
smoothly with $\Delta _{0}$ (i.e. $N_{s}$ decreases slowly). The vanishing
of the cloud through a narrow bistable cycle is not visible on the graph, as
it occurs closer from resonance. As $n_{0}$ increases, the bistable cycle
appears for smaller $Z_{s}$ (and thus larger $N_{s}$), and becomes
physically significant. Fig. \ref{fig:foldex}b shows $Z_{s}$ for $n_{0}$\ $%
=2.5\times 10^{8}$ and a bistable cycle for $-0.3\lesssim \Delta
_{0}\lesssim -0.25$. If $n_{0}$ is further increased, the bistable cycle
disappears, but it remains a fold corresponding to two abrupt slope changes
of $Z_{s}$ versus $\Delta _{0}$ (fig. \ref{fig:foldex}c, $n_{0}$\ $%
=3.4\times 10^{8}$). If $n_{0}$ is still increased, the fold remains, but it
becomes smoother (Fig. \ref{fig:foldex}d for $n_{0}=$\ $4\times 10^{8}$).

The dynamics of the cloud is determined by the stability of these stationary
solutions. In particular, if no stationary solutions are stable, complex
dynamics could be obtained. The stability of the above stationary solutions
is evaluated through a linear stability analysis, which associates to each
stationary solution its three eigenvalues, corresponding to the stability
following its three eigendirections in the 3D phase space of our model. The
real part of the eigenvalue, corresponding to a damping rate, determines the
stability (stable if negative). The imaginary part, when different from
zero, is associated to an angular eigenfrequency, also called relaxation
frequency, which play a main role in the dynamics. A pleasant -- and simple
-- way to describe the stationary solutions is to use their phase space
representation, where each stationary solution corresponds to a fixed point
with its properties depending on its eigenvalues. Let us remember that the
standard terminology distinguishes the stable node (all eigenvalues real and
negative), the stable focus (all real parts negative, two eigenvalues
complex conjugate), the saddle node (all eigenvalues real, at least one
positive) and the saddle focus (at least one real part positive, two
eigenvalues complex conjugate). For sake of simplicity, this terminology
will be used in the following.

The set of points of vertical tangency in $Z_{s}$ determines a line which
delimits the unstable stationary solution of the medium branch of the
bistability cycle. Linear stability analysis shows that the fixed point in
that area is a saddle-node with real eigenvalues, two being positive and one
negative. But the unstable stationary solutions extend beyond this area. In
particular, the stationary solutions can also be unstable on a part of the
upper branch of the bistable cycle, and even outside the bistable cycle,
when the stationary solution is unique. This is illustrated on fig. \ref
{fig:ZS} where the nature of the stationary solution is indicated by a level
of gray and a code. In the SN zone, the fixed point associated with the
stationary solution is a saddle node, i.e. the stationary solution is
unstable, and its three eigenvalues are real, two being positive. In the SF
zone, the fixed point is a saddle focus, i.e. the stationary solution is
unstable, and two of its eigenvalues are complex with a positive real part,
and the third one is real negative. In the F zone, the fixed point is a
stable focus, i.e. the stationary solution is stable, and two of its
eigenvalues are complex with a negative real part, the third one is real
negative. Finally, in the N zone, the fixed point is a stable node, i.e. the
stationary solution is stable and all eigenvalues are real negative. The
dashed line indicates the location of the bifurcation, i.e. the transition
from stable to unstable stationary solutions. In most cases, it occurs from
stable focus to saddle focus, through a super-critical Hopf bifurcation.

As the detuning is varied, four typical situations may occur, already
illustrated in Fig. \ref{fig:foldex}. For small $n_{0}$ (Fig. \ref
{fig:foldex}a), $Z_{s}$ is always stable, and its dependence versus $\Delta
_{0}$ is almost flat (except very close from resonance): we expect a
stationary cloud slightly moving with the detuning. In the bistability area
(Fig. \ref{fig:foldex}b), the central branch of the bistability cycle is
unstable, as usual in such a situation, but the upper branch is also partly
unstable, reducing drastically the parameter range where the system is
effectively bistable. However, a narrow bistable zone should be observed. As 
$n_{0}$ is still increased, only one stationary solution remains (Figs \ref
{fig:foldex}c), which is unstable on the fold.

\begin{figure}[tph]
\centerline{\resizebox{0.45\textwidth}{!}{\includegraphics{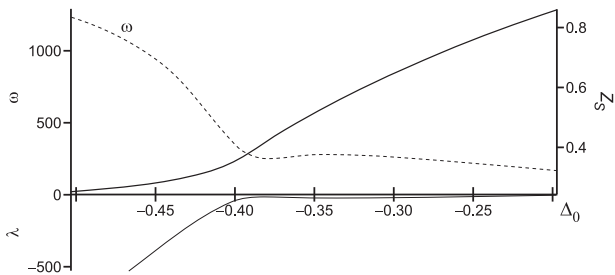}}}
\caption{Evolution as a function of the detuning of the stationary solution $%
Z_S$ and its eigenvalues, for the parameters of Fig. \ref{fig:foldex}d. The
stationary solution is given through the full bold line in the upper side of
the figure. The dashed line noted $\protect\omega$ represents the imaginary
part of the complex eigenvalues, while the full line corresponds to their
real part $\protect\lambda$. The third eigenvalue is always real negative,
almost constant and everywhere larger than $\protect\lambda$. It cannot be
distinguished from the zero axis at the scale of the figure.}
\label{fig:foldstab1}
\end{figure}

\begin{figure}[tph]
\centerline{\resizebox{0.45\textwidth}{!}{\includegraphics{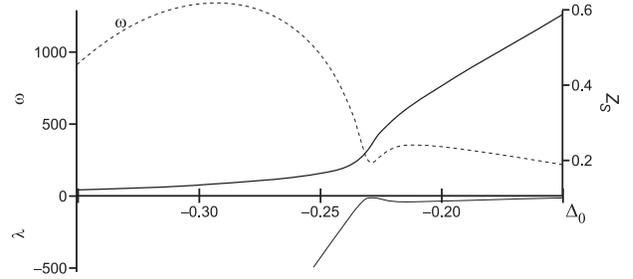}}}
\caption{Same as Fig. \ref{fig:foldstab1} for parameter set \#2 of Table \ref
{tabtheoparams}. For sake of clarity, the evolution of $\protect\lambda$ has
been truncated for small detunings, where it becomes almost linear, reaching
e.g. the value of $\protect\lambda=-1500$ for $\Delta_0=-0.3$.}
\label{fig:foldstab3}
\end{figure}

Finally, for large $n_{0}$ (Fig. \ref{fig:foldex}d), the solution is always
stable (F zone), except very close from resonance. Thus we expect to observe
a stationary cloud moving with $\Delta _{0}$. However, because of the fold,
this moving is non linear, and on the fold, for $-0.45<$ $\Delta _{0}<-0.35$%
, $Z_{S}$ is more sensitive to the $\Delta _{0}$ value, because the slope
here is larger. On the other hand, a closer look to the eigenvalues (Fig. 
\ref{fig:foldstab1}) shows that at the level of the fold, the real part $%
\lambda $ of the complex eigenvalues $\lambda \pm i\omega $ is close to
zero, while the eigenfrequency crosses a minimum. This is more evident on
Fig. \ref{fig:foldstab3}, where the same situation is illustrated for a
different set of parameters. The vanishing damping rate $\left| \lambda
\right| $ associated to an increased sensitivity to the parameter values
means that a small perturbation will cause a large undamped reaction of the
system, with relaxation oscillations at a frequency $\omega $ crossing a
minimum while $\Delta _{0}$ is varied on the fold. In the fold inflection
point, where the slope is maximum, the eigenfrequency and damping rate are
minimum: the effects of perturbations is expected to be maximum at this
point. Note that the evolution of $\lambda $ and $\omega $ around the
inflection point is asymmetric: the damping rate and relaxation frequency
increase rapidly for detunings smaller than the inflection point, while they
remain of the same order of magnitude on the other side. This asymmetry is
expected to have consequences on the dynamics.

Deterministic instabilities are expected to occur for parameters where all
stationary solutions are unstable, i.e. in the SF monostable zone\cite
{InstDet}. In all other areas, the stationary solutions are stable, and
therefore, deterministic instabilities cannot occur. However, the presence
of the fold and the proximity of a Hopf bifurcation generate a type of
stochastic behavior similar to instabilities\cite{nousprl}.

Indeed, on the one hand, the proximity of a Hopf bifurcation is known to
favor the appearance of coherence resonance\cite{SRlike}, as discussed in 
\cite{nousprl}. Let us recall that coherence resonance is the counterpart of
stochastic resonance in autonomous systems\cite{SRRevue}. Stochastic
resonance appears in some forced systems: it may be seen as an amplification
by noise of the system response to a forcing. In other terms, the signal to
noise ratio of the output periodic signal resulting from the modulation
exhibits a maximum when the noise amplitude increases. In stable autonomous
systems, the behavior is not periodic, but a phenomenon similar to
stochastic resonance can lead to an amplification of an internal resonance:
it is the internal stochastic resonance, or coherent resonance \cite
{CohRes,CohResPrecursor}. Coherent resonance is also known to generate noise
induced coherent oscillations, which explain some of our experimental
observations, as shown in \cite{nousprl}.

On the other hand, as shown above, the particular configuration of the
eigenvalues on the stable fold makes the system\ very sensitive to small
perturbations, i.e. to noise, and thus we expect a large increasing of the
response to noise on the fold. The resulting behavior looks like
instabilities, whereas it is simply noise amplification. The behavior
obtained on the stable fold when noise is taken into account is detailed in
the next section.

\section{The stable fold: stochastic instabilities}

\label{S7StaFold}In \cite{InstDet}, it has been shown that the model is able
to reproduce the {\it C} instabilities, in the monostable SF zone. On the
contrary, {\it S} instabilities do not appear in this situation. In fact, it
has been shown in \cite{nousprl} that {\it S} instabilities are not
instabilities in the usual significance, but the result of the amplification
of the system intrinsic noise, and thus it is necessary to add noise in the
simulations to observe {\it S} instabilities. This dynamics could appear
when the stationary solution of the MOT is not stable, but would be
difficult to analyze, because in this case the resulting dynamics would be
the superimposition of the deterministic instability with the stochastic
motion, the latter being masked by the former. For this reason, the present
section is devoted to the study of the influence of noise on the stable
stationary solutions of the model, in particular in the vicinity of the
fold. As the model here is slightly different from that used in \cite
{nousprl}, we could expect different results from those obtained in \cite
{nousprl}. It is shown in the following that the modifications introduced in
the model do not alter the previous conclusions, i.e. that {\it S}
instabilities are produced by noise amplification on the fold.

\begin{figure}[tph]
\centerline{\resizebox{0.45\textwidth}{!}{\includegraphics{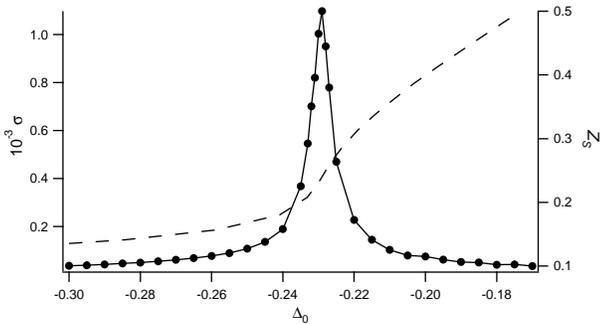}}}
\caption{Evolution of the signal standard deviation $\protect\sigma$ (full
line) and of the cloud location stationary solution $Z_S$ (dashed line) as a
function of the detuning $\Delta_0$. White noise is applied to $I_1$ with $%
\protect\zeta=10^{-3}$. Other parameters correspond to the set \#2 of Tab. 
\ref{tabtheoparams}. }
\label{fig:theomax}
\end{figure}

To evaluate the response of the system to noise in the vicinity of the fold,
we have plotted the amplitude of the motion perturbed by noise, versus the
detuning, across the fold (Fig. \ref{fig:theomax}). The amplitude of the
motion is measured through the standard deviation $\sigma $ of $Z$. Fig. \ref
{fig:theomax} has been obtained for a gaussian white noise applied on $I_{1}$
with an amplitude of $10^{-3}$, in a situation where the stationary
solutions are stable on the fold. Noise amplification appears clearly on the
fold, and mimics instabilities, as in the experiments (Fig. \ref{fig:expmax}%
). The maximum of the motion amplitude, obtained for $\Delta _{0}=-0.23$,
corresponds to a standard deviation of $1.1\times 10^{-3}$, i.e. $33$~$\mu $%
m, in good agreement with the experimental value. By comparison, the motion
amplitude is $0.036$ for $\Delta _{0}=-3$, i.e. the effect of noise is $30$
times larger on the fold than in $\Delta _{0}=-3$. Note that, as similar
motion amplitudes are obtained in the experiment and in the model for a
noise level $\zeta =10^{-3}$, this gives us in return an evaluation of the
experimental noise level.

Thus, as for the model developed in \cite{nousprl}, the fold plays the role
of a noise amplifier, leading to an unstable cloud over a limited range of
the parameters. The maximum of the motion amplitude appears at the fold
inflection point. The amplitude decreases progressively and quasi
symmetrically on each side of this point, in spite of the asymmetry of the
eigenvalues discussed in the previous section. The range where the
instabilities can be observed, appears to be of the order of the ``width''
of the fold, that we could define as the interval between the two abrupt
slope changes delimiting the fold (a more precise definition could be given,
but is not useful for the following). In the case discussed here, the range
in detuning is 0.02, i.e. much smaller than the experimental range, which
was typically of the order of 1. However, the width of the slope is very
dependent on the MOT parameters: for example, on Fig. \ref{fig:foldstab1},
where $I_{1}$ and $n_{0}$ are different, the fold is more than twice larger
than in the present case. Thus there is no doubt that it is possible to find
a parameter set giving a correct fold width. However, because of the extreme
simplicity of the model as compared to the experiments, such a search has no
meaning: the aim here is just to show that very few ingredients, including
the shadow effect and the noise, are able to reproduce the global
experimental behavior. We have now demonstrated that the noise response can
effectively have the appearance of instabilities for some parameters. We
will now examine in detail the resulting dynamics, and compare it to the
experimental observations.

\begin{figure}[tph]
\centerline{\resizebox{0.45\textwidth}{!}{\includegraphics{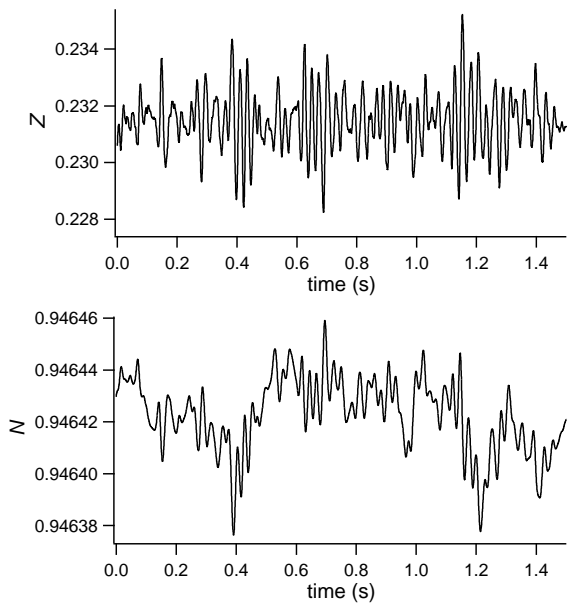}}}
\caption{Dynamics obtained by resolving Eqs. (\ref{eqred}), with the
parameter set \#2 and $\Delta_0=-0.926 $. White noise is applied to $I_1$
with $\protect\zeta=10^{-3}$.}
\label{fig:TheoTimeS}
\end{figure}

\begin{figure}[tph]
\centerline{\resizebox{0.45\textwidth}{!}{\includegraphics{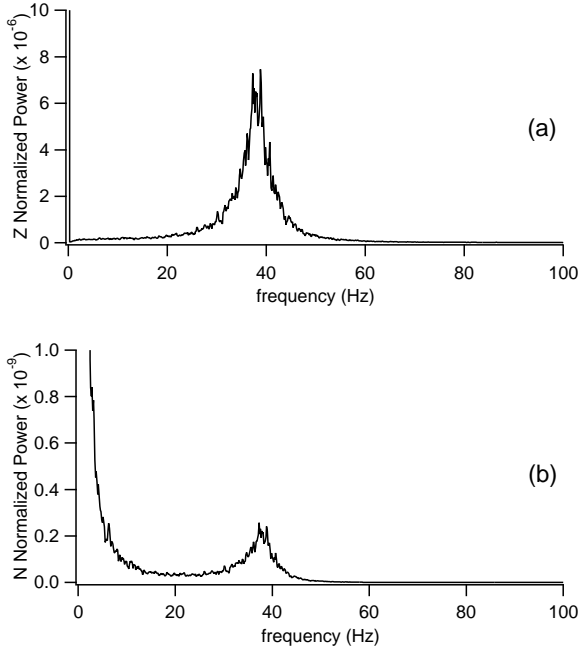}}}
\caption{Power spectra obtained by resolving Eqs. \ref{eqred} with the same
parameters as in Fig. \ref{fig:TheoTimeS}.}
\label{fig:TheoSpectreS}
\end{figure}

Fig. \ref{fig:TheoTimeS} shows the time evolution of $Z$ and $N$ in $\Delta
_{0}=-0.23$, where the amplification is maximum. The signal is of course
stochastic, but several differences appear, on the one hand between the
cloud response and the applied noise, and on the other hand between the $Z$
and $N$ evolutions. Indeed, while the applied noise is white, a frequency
dominates in the response, in particular for $Z$. This is clearly visible on
the signal, and of course on the spectrum (Fig. \ref{fig:TheoSpectreS}),
where a peak appears at a frequency $\omega _{S}\simeq 38$ Hz. For $N$,
while the frequency $\omega _{S}$ still appears, low frequency fluctuations
dominates, as it is also shown by the spectrum (Fig. \ref{fig:TheoSpectreS}%
). This behavior with two different characteristic times is very similar to
the experimental {\it S}$_{{\it H}}$ behavior illustrated in the Fig. \ref
{fig:exp2freq}: in both cases, the higher frequency component dominates the
motion of the atoms, while the number of atoms in the cloud is mainly driven
by the low frequency component.

\begin{figure}[tph]
\centerline{\resizebox{0.45\textwidth}{!}{\includegraphics{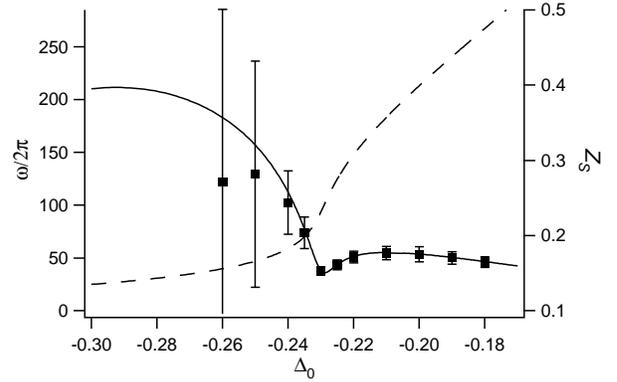}}}
\caption{Evolution of the resonance frequencies as a function of the
detuning. The full and dashed line recalls the evolution of $\protect\omega$
and $Z_S$, already shown on Fig. \ref{fig:foldstab3}. The squares represent
the resonance frequencies appearing in the dynamics, obtained by fitting a
lorentzian function on spectra similar to those of fig. \ref
{fig:TheoSpectreS}. The bars give the width of the lorentzian.}
\label{fig:freqres}
\end{figure}

In the experiments, the origin of the higher frequency was not identified,
as it does not correspond to a known experimental characteristic frequency.
However, we know now that the stable stationary solution on the fold has
complex eigenvalues, and thus is associated with a relaxation frequency. As
noise is known to be able to excite such non linear resonance
eigenfrequencies\cite{celet}, it is interesting to compare in the
simulations the frequency appearing in the dynamics with the eigenfrequency
of the stationary solution. This is illustrated on Fig. \ref{fig:freqres}
for different values of the detuning. The full and dashed lines concern the
stationary solutions: they shows the evolution of $\omega $ and $Z_{S}$
versus $\Delta _{0}$. They are a reproduction of Fig. \ref{fig:foldstab3},
and are recalled just for comparison. The squares give the resonance
frequencies of the noisy dynamics. They are obtained by fitting \ the
calculated spectra of $Z$, as illustrated on Fig. \ref{fig:TheoSpectreS}a,
with a lorentzian function. Similar fits on the $N$ spectra give the same
results. Fig. \ref{fig:freqres} shows that for detunings larger than the
fold inflection point, the noisy resonance frequency and the eigenfrequency
correspond exactly. This confirms that the frequency appearing in the
dynamics is the relaxation frequency appearing in the eigenvalues associated
with the stationary solution. Thus ``instabilities'' appear in fact as a
noisy amplification of the relaxation frequency of the MOT, through a
phenomenon already observed in many other systems, as e.g. in lasers \cite
{celet}: the small damping rate allows noise to excite the relaxation
frequency, altering the frequency distribution of the system response to
noise.

For detunings smaller than the fold inflection point, Fig. \ref{fig:freqres}
exhibits differences between the eigenfrequencies and the resonance
frequencies. A close look at the dynamics of the cloud shows that the main
point is not this difference, but a dramatic broadening of the noise
resonance, leading to an indetermination of the resonant frequency. This is
illustrated on Fig. \ref{fig:freqres}, where the bars associated with each
point have a height of $\Delta \nu $, the width of the lorentzian peak
obtained by fitting the $Z$ spectrum. On the inflection point, the resonance
is narrow, as it appears on Fig. \ref{fig:TheoSpectreS}. On the right of the
inflection point, the resonance remains narrow, reaching a maximum of $%
\Delta \nu =14$ Hz in $\Delta _{0}=-0.2$. On the contrary, for detunings
smaller than the inflection point, $\Delta \nu $ increases rapidly, reaching
already the value $\Delta \nu =30$ Hz in $\Delta _{0}=-0.235$ (the
inflection point is in $\Delta _{0}=-0.23$). It is clear on Fig. \ref
{fig:freqres} that the discrepancy between the noise resonance and the
relaxation frequencies is linked to the width of the resonance: as the
resonance broadens, it also flattens, and the central frequency becomes
irrelevant. This is the reason why the resonant frequencies for detunings
smaller than $\Delta _{0}=-0.26$ are not reported on the figure.

\begin{figure}[tph]
\centerline{\resizebox{0.45\textwidth}{!}{\includegraphics{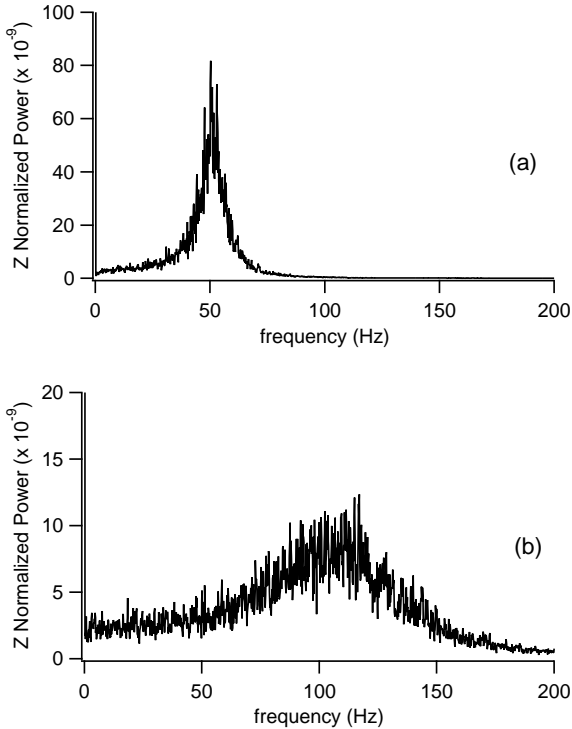}}}
\caption{$Z$ spectra of the instabilities for two values of the detuning
situated symmetricallty with respect to the fold maximum. In (a), $%
\Delta_0=-0.22$; in (b), $\Delta_0=-0.24$. Note that the vertical scale in
(a) is five times larger than in (b).}
\label{fig:asym}
\end{figure}

This asymmetry in the dynamics around the fold inflection point is of course
linked to the asymmetry in the eigenvalues associated with the stationary
solution. For detunings larger than the inflection point, the damping rate
remains small (less than 45 s$^{-1}$), and thus the noise excitation of the
relaxation oscillations remains efficient. On the contrary, for detunings
smaller than the inflection point, the damping rate increases rapidly,
decreasing the efficiency of the excitation, and leading to a flat
resonance. The consequence on the dynamics is illustrated on Fig. \ref
{fig:asym}, where the power spectra of the $Z$ dynamics are represented for
two different values of the detuning, located symmetrically with respect to
the inflection point. In (a), for $\Delta _{0}=-0.22$, i.e. for a larger
detuning than the inflection point, the resonance, centered on 50 Hz,
remains narrow ($\Delta \nu =11$ Hz), and thus significant. In (b), for $%
\Delta _{0}=-0.24$, i.e. on the other side of the inflection point, the
resonance, centered on 100 Hz, is already 60 Hz wide, and also five times
lower (note the different vertical scales in (a) and (b)). Such a resonance
is no more significant from an experimental point of view: indeed, the
experimental dynamics is analyzed from time series with a necessary limited
number of points, leading to spectra with a resolution much smaller than in
the simulations. It is clear that a resonance as those observed on the left
of the inflection point could not be detected in the experiments, and thus
one expects that in this situation, experiments deliver non resolved
spectra. It was effectively the case for {\it S}$_{{\it L}}$ instabilities
(Fig. \ref{fig:exp1freq}), and thus we can conclude that the {\it S}$_{{\it L%
}}$ dynamics observed in the experiments is also explained by the present
model.

The results obtained in this section can be summarized in the following way:
when the parameters of the cloud are such that the unique stationary
solution is stable, and therefore that no deterministic instabilities can
occur, the action of noise mimics instabilities. This originates in a fold
of the stationary solution, which plays the role of a noise amplifier.
Moreover, the particular properties of the eigenvalues associated with the
stationary solutions on the fold -- small damping rate and relaxation
frequency -- lead to the existence of the stochastic regime: for detuning
larger than the fold inflection point, a resonance is excited by the noise,
leading to a dynamics with a dominating frequency; for detuning smaller than
the inflection point, the damping rate increases, and the resonance
vanishes. This allows us to interpret all the regimes observed
experimentally.

The particular properties of the eigenvalues discussed above are in fact
linked to the proximity of a Hopf bifurcation. Another\ known consequence of
such a situation is to favor the existence of coherence resonance, and such
a phenomenon exists effectively in our case \cite{nousprl}. Now, an
interesting question is to determine what are the exact connections between
all these phenomena, and in particular if the existence of coherence
resonance has a determining role in that of instabilities. The answer is
obviously: no! Indeed, the essential ingredient for noise amplification is
the fold; the properties of the eigenvalues only determine the time
characteristics of the dynamics. In particular, we have checked that with
the present model, no coherence resonance occurs when the MOT exhibits {\it S%
}$_{{\it L}}$ instabilities.

\section{Conclusion}

It has been shown recently that the cloud of cold atoms obtained from a
magneto-optical trap may exhibit a complex dynamics in the regime of high
atomic densities. The observed behaviors can be essentially separated in two
different types, depending of their nature: stochastic instabilities \cite
{nousprl} or deterministic instabilities\cite{InstDet}. The aim of this
paper was to describe extensively the experimentally observed stochastic
dynamics, and to understand its mechanisms, through a model briefly
presented in \cite{InstDet} and detailed here.

A detailed analysis of the experimental results in the stochastic regime
shows a variety of dynamical behaviors, which differ by the frequency
components appearing in the dynamics. Indeed, some instabilities exhibit
only low frequency components, while in other cases, a second time scale,
corresponding to a higher frequency, appears in the motion of the center of
mass of the cloud.

The simple stochastic 1D-model that we use here allows us to retrieve and
interpret these experimental dynamics. The model shows that the existence of
instabilities is linked to folded stationary solutions where noise response
is enhanced. Moreover, the proximity of a Hopf bifurcation and the resulting
conditions on the stability of the stationary solutions -- small damping
rate and existence of a relaxation frequency -- explains the existence of
several types of regimes: indeed, depending on the parameters, noise is
sometimes able to excite the relaxation frequency, leading to the appearance
of the second time scale in the dynamics. Globally, the agreement between
this over-simplified 1D model and the 3D experiments is surprisingly good.
Not only it allows us to make a qualitative interpretation of the
experimentally observed dynamics, but also gives quantitative results
compatible with the experimental measures, in particular concerning the
amplitude of the instabilities. However, it is clear that a 3D model is
necessary for an accurate quantitative description of all the experimental
observations.

The model also emphasized the close relations between the stochastic and
deterministic instabilities. Indeed, both types of dynamics appear to be
link to the same factors: the effective dynamics depends mainly on the
distance between the working point and the bistable cycles (or the Hopf
bifurcation). In fact, the dynamics described in the present paper appears
as the first sign of the deterministic instabilities described in \cite
{InstDet}, and thus may be considered as a noisy precursor to deterministic
instabilities\cite{paramaccess,precursor}. A study of the phenomenon from
this point of view should put in evidence new properties of these
instabilities.

\section{Acknowledgments}

The author thanks M. Fauquembergue and A. di Stefano for their participation
in the elaboration of the model, D. Wilkowski for its contribution to the
first stages of the experiments, and Ph. Verkerk for useful discussions and
his comments about this paper. The Laboratoire de Physique des Lasers,
Atomes et Mol\'{e}cules is ``Unit\'{e} Mixte de Recherche de l'Universit\'{e}
de Lille 1 et du CNRS'' (UMR 8523). The Centre d'Etudes et de Recherches
Lasers et Applications (CERLA) is supported by the Minist\`{e}re charg\'{e}
de la Recherche, the R\'{e}gion Nord-Pas de Calais and the Fonds Europ\'{e}%
en de D\'{e}veloppement Economique des R\'{e}gions.

% now the references. delete or change fake bibitem. delete next three
% lines and directly read in your .bbl file if you use bibtex.

\end{document}